# Model of the Electrostatics and Tunneling Current of Metal-Graphene Junctions and Metal-Insulator-Graphene Heterostructures

Ferney A. Chaves[1], David Jiménez[1], Aron W. Cummings[2], and Stephan Roche[2,3]

[1]*Departament d'Enginyeria Electrònica, Escola d'Enginyeria, Universitat Autònoma de Barcelona, Campus UAB, 08193 Bellaterra, Spain.*
[2]*ICN2 – Institut Catala de Nanociencia i Nanotecnologia, Campus UAB, 08193 Bellaterra (Barcelona), Spain.*
[3]*ICREA, Institució Catalana de Recerca i Estudis Avançats, 08070 Barcelona, Spain.*

In this paper we present a comprehensive model for the tunneling current of the metal-insulator-graphene heterostructure, based on the Bardeen Transfer Hamiltonian method, of the metal-insulator-graphene heterostructure. As a particular case we have studied the metal-graphene junction, unveiling the role played by different electrical and physical parameters in determining the differential contact resistance.

*Introduction*. The metal-graphene (MG) junction is a critical component of graphene-based devices and controlling its properties is a pre-requisite for device optimization. Although there are quite a few experimental studies of the MG junction in the scientific literature [1-5], a comprehensive model of the tunneling current between a pure 2D material and a metal is still lacking. These kinds of models are needed, for instance, to get information about what can be done to optimize the contact resistance. Although the carrier mobility in the graphene is very high, its small density of states (DOS) might suppress current injection, limiting the performance of graphene-based devices [6].

We address this problem by formulating an analytical model of the metal-insulator-graphene (MIG) heterostructure, from which the MG junction can be seen as a particular case. Our model is based on the Bardeen Transfer Hamiltonian (BTH) method [7,8] where the probability of elastic tunneling is calculated using Fermi golden rule. This gives a quantitative estimate of the coupling between metal and graphene states, allowing us to obtain an analytical formula for the tunneling current, which contains key information on the role played by different parameters.

The manuscript is organized as follows: in Section I we address the electrostatics of the MIG heterostructure. Specifically, a simple model to calculate the location of the Fermi level in both electrodes is presented. In Section II, we present a model for the tunneling current. In Section III some results are presented in terms of benchmarking different metal electrodes. Finally, we outline the main conclusions of this work.

**I. *Electrostatics of the MIG heterostructure*.** The MIG heterostructure, represented in Fig. 1, consists of metal (M) and graphene (G) electrodes with work functions $W_m$ and $W_g$, respectively, separated by an intermediate insulator (I) layer. The (I) layer can represents either a dipole layer formed as a result of charge transfer within the equilibrium separation distance $d = d_{eq}$ with $\varepsilon = \varepsilon_0$ (case of MG junction), or a general dielectric of permittivity $\varepsilon > \varepsilon_0$ and thickness $d$ (case of MIG heterostructure).

After formation of the MIG structure a charge transfer from (G) to (M) is produced until their Fermi levels align Figs. 1b,c. A dipole is formed at the (I) layer with a potential drop $\Delta V$ across it and the graphene becomes doped because a shift $\Delta E = E_{fg} - E_D$ of its Fermi level ($E_{fg}$) with respect to the Dirac point ($E_D$). The value of $\Delta E$ depends on metal nature and the ratio $z_d/\varepsilon$, with $z_d$ being the effective distance between the charge sheets of (M) and (G) electrodes [9]. In this work, we model it as $z_d = d - d_0$ with constant $d_0 \sim 0.24\ nm$.

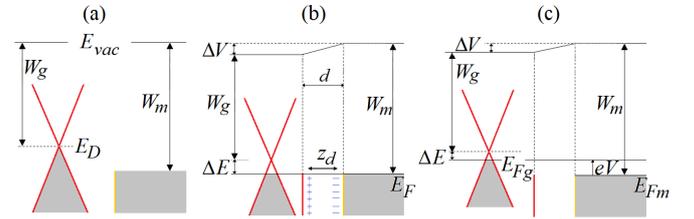

FIG. 1. (a) Band diagram of an isolated graphene and metal system. (b) Band diagram of a MG junction at equilibrium showing dipole formation at the interface. A voltage drop $\Delta V$ is produced over the dipole, and $\Delta E$ represents the shift of the graphene Fermi level with respect to the Dirac point. (c) Non-equilibrium band diagram of the MG junction.

When a bias voltage $V$ is applied between the two electrodes, $\Delta E$ changes according to the relation

$$\Delta E = W_m - W_g - e\Delta V - eV. \quad (1)$$

The potential drop $\Delta V$ can be expressed as $\Delta V = \Delta_{tr} + \Delta_{ch}$, where the $\Delta_{tr}$ term results from charge transfer and the $\Delta_{ch}$ term describes the short-range interaction from the overlap of wavefunctions, which depends strongly on the separation distance $d$ and it becomes negligible for $d \gtrsim 5$nm [9]. At $V = 0$ the doped type of the graphene in a MIG structure, in general, is determined by $eV_D = W_m - W_g - e\Delta_{ch}$. To model the electron transfer contribution, $\Delta_{tr}$, we use a planar capacitor model giving $\Delta_{tr}(\Delta E) = ez_d(n-p)/\varepsilon$, where $n - p$ is the net 2D carrier density in the graphene, and $e$ is the electron charge. Using the usual expression for $n - p$ [10] into Eq. (1) yields the relation

$$\alpha f(\Delta E/k_B T) + \Delta E + eV - eV_D = 0 \quad (2)$$

where $\alpha = 2e^2 z_d/(\varepsilon \pi \hbar^2 v_f^2)$ and $f(x) = (k_B T)^2 [\mathcal{F}_1(-x) - \mathcal{F}_1(x)]$, with $\mathcal{F}_1(x)$ the Fermi-Dirac integral of order 1.

The current-voltage (I-V) characteristic of a MIG junction can be understood by considering the possible locations of the Fermi levels around the graphene Dirac point, as illustrated in Fig. 2. The bias voltage $V$ changes the relative difference between the Fermi levels on each side according to $E_{fg} - E_{fm} = eV$. We assume that $\Delta E \geq 0$ when $E_{fg} \leq E_D$ and $\Delta E < 0$ otherwise. Depending on the bias voltage $V$, several regions of operation arise. If $V > 0$ a positive current will flow from the graphene to the metal via tunneling across the (I) layer as shown in Figs. 2a-c. At $V = V_D$, a perfect alignment of the graphene Fermi level with the Dirac point is produced (Fig. 2b), resulting in $\Delta E = 0$. On the other hand, if $V < 0$, a negative tunneling current flows across the (I) layer (Figs. 2d-f). In the latter case $V_c$ is the bias voltage needed to align the $E_{fm}$ level to the Dirac point, such that $eV = -\Delta E$.

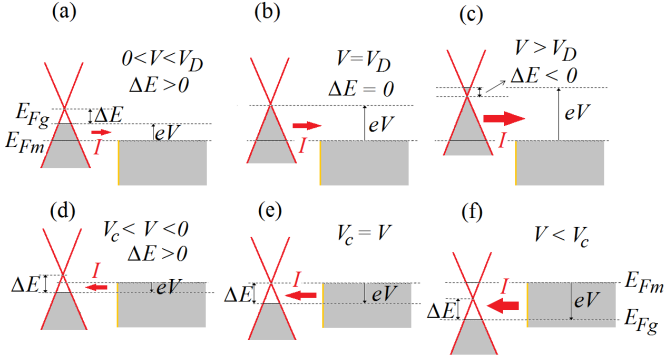

FIG. 2. Schematic representation of the tunneling current between graphene and metal electrodes. For panels (a)-(c) the current is due to electron flow from the graphene into the metal. For panels (d)-(e) the current flows from the metal into the graphene. Note that the Dirac point and the graphene Fermi level vary their positions with the bias voltage.

**II. *Tunneling in a MIG heterostructure.*** Using the BTH approach [7,8] the tunneling current is computed via the expression

$$I = g_V \frac{4\pi e}{\hbar} \sum_{g,m} |M_{g\leftrightarrow m}|^2 [f_g(E_g) - f_m(E_m)] \delta(E_g - E_m) \quad (3)$$

where $g$ and $m$ label the states in the (G) and (M) electrodes with energies $E_g = E_g(\mathbf{k}_g) = E_D \pm \hbar v_f |\mathbf{k}_g|$ and $E_m = E_m(\mathbf{k}_m, k_z) = \hbar^2(k_m^2 + k_z^2)/2m$, respectively, which we have sketched in Fig. 3 for convenience. Here $g_V$ is the electron valley degeneracy and $f_g$ and $f_m$ are the electron Fermi occupation factors. The $M_{g\leftrightarrow m}$ term refers to the matrix element for the transitions and it is given by

$$M_{g\leftrightarrow m} = \frac{\hbar^2}{2m_0} \iint \left( \Psi_g^* \frac{d\Psi_m}{dz} - \Psi_m \frac{d\Psi_g^*}{dz} \right) dS, \quad (4)$$

where $m_0$ the electron mass in the (I) layer. The $\Psi_g(\mathbf{r}, z)$ and $\Psi_m(\mathbf{r}, z)$ represent electron wavefunctions. Considering the graphene with two identical atoms per unit cell, labeled 1 and 2, the wavefunction for wavevector $\mathbf{k}$ can be written in terms of the basis functions $\Phi_{j\mathbf{k}}$ ($j = 1,2$) on each atom as $\Psi_g(\mathbf{r}, z) = \sum_j \chi_j(\mathbf{k}_g) \Phi_{j\mathbf{k}_g}(\mathbf{r}, z)$. The basis functions have Bloch form, $\Phi_{j\mathbf{k}_g}(\mathbf{r}, z) = \exp(i\mathbf{k}_g \cdot \mathbf{r}) u_{j\mathbf{k}_g}(\mathbf{r}, z)/\sqrt{A}$, where $u_{j\mathbf{k}_g}(\mathbf{r}, z)$ is a periodic function and $A$ refers to the contact area. These periodic functions are localized around the basis atoms (i.e., as 2p$_z$ orbitals) of the graphene, and $u_{j\mathbf{k}_g}(\mathbf{r}, z)$ is expected to vary only weakly along the radial coordinate $\mathbf{r}$ in the graphene. Thus, we assume that $u_{j\mathbf{k}_g}(\mathbf{r}, z) = f_{j\mathbf{k}_g}(\mathbf{r})g(z)$ and we approximate the radially-dependent term $f_{j\mathbf{k}_g}(\mathbf{r})$ as numerical constants $f_1$ and $f_2$ [10]. The z-dependence has the usual decaying form $g(z) = e^{-\kappa z}/\sqrt{D}$, where $\kappa$ is the decay constant of the wavefunction in the barrier, and $D$ is the normalization constant. Both $\chi_1(\mathbf{k}_g)$ and $\chi_2(\mathbf{k}_g)$ have well-known values for graphene in a nearest-neighbor tight-binding approximation [11].

On the other hand, the metal electrons can be modeled as free incident and reflected particles for $z \geq d$ and with a decaying exponential for $z < d$, namely

$$\Psi_m(\mathbf{r}, z) = \begin{cases} \frac{e^{i\mathbf{k}_m \cdot \mathbf{r}}}{\sqrt{A}} * \frac{te^{\kappa(z-d)}}{\sqrt{D}}, & z < d \\ \frac{e^{i\mathbf{k}_m \cdot \mathbf{r}}}{\sqrt{A}} * \frac{1}{\sqrt{L}} [e^{-ik_z(z-d)} + re^{ik_z(z-d)}], & z \geq d \end{cases} \quad (5)$$

where $t$ and $r$ are the amplitudes of the transmitted and reflected waves, respectively, and $L$ is a normalization constant. As usual, the wavefunctions matching conditions in $z = d$ have to be fulfilled, resulting in $t = 2\sqrt{D/L}k_z/(k_z + i\kappa)$. Thus, the square of the matrix elements can be written as

$$|M_{g\leftrightarrow m}|^2 \approx \left(\frac{\hbar^2}{2m_0} \frac{4\kappa e^{-\kappa d}}{\sqrt{D}}\right)^2 |\Theta|^2 \omega(k_z) \frac{1}{L} \left|\frac{1}{A} \int dS\, e^{i(\mathbf{k}_g - \mathbf{k}_m)\cdot \mathbf{r}}\right|^2, \quad (6)$$

where we have defined the functions $\Theta(\theta_{\mathbf{k}_g}) = [\chi_1^* f_1^* + \chi_2^* f_2^*]$ and $\omega(k_z) = k_z^2/(k_z^2 + \kappa^2)$. The integral on the right-hand side of Eq.(6) approaches the delta-function $\delta(\mathbf{k}_g - \mathbf{k}_m)$ when $A \to \infty$, implying the conservation of in-plane momentum $\mathbf{k}$. Incorporating Eq. (6) into Eq. (3), we get the following expression for the tunneling current

$$I \propto \sum_{\mathbf{k}_g, \mathbf{k}_m, k_z} |\Theta(\theta)|^2 \omega(k_z) [f_g(E_g) - f_m(E_m)] \delta(E_g - E_m) \delta_{\mathbf{k}_g, \mathbf{k}_m}. \quad (7)$$

The delta-function in Eq. (7) guarantees that only processes conserving the energy are possible. In the limit of large $A$ we have $\mathbf{k}_g = \mathbf{k}_m \equiv \mathbf{k}$ where $|\mathbf{k}| = k$, and converting the discrete sums to integrals, the equation for the tunneling current becomes

$$J = \eta(\kappa) \frac{2m}{\hbar^2} \iint dk_z dk k \omega(k_z)(f_g - f_m) \delta[(k - k_1)(k - k_2)], \quad (8)$$

where $\eta(\kappa) = \frac{8\pi e}{\hbar} \left(\frac{\hbar^2}{2m_0} \frac{4\kappa e^{-\kappa d}}{\sqrt{D}}\right)^2 \frac{|f_1|^2}{(2\pi)^2}$ with $|f_1|^2$ constant of order unity assumed to have no dependence on $\mathbf{k}$. The values $k_1$ and $k_2$ are the solution of the quadratic equation $E_g(k) - E_m(k) = 0$ given by

$$k_1 = \xi + \sqrt{\xi^2 + k_D^2 - k_z^2}, \quad k_2 = \mp\xi \pm \sqrt{\xi^2 + k_D^2 - k_z^2}. \quad (9)$$

The upper (lower) sign of $k_2$ applies to the valence (conduction) band where we have defined the constants $\xi = mv_f/\hbar$ and $k_D^2 = 2m/\hbar^2 E_D$. Physically, $k_1$ and $k_2$ are the in-plane momentum $k$ values satisfying both the in-plane momentum and the energy conservation conditions (Fig. 3).

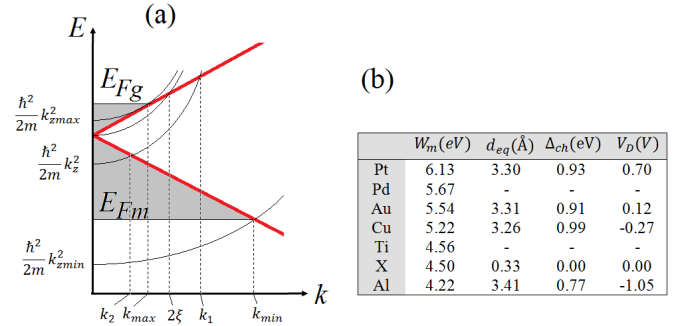

FIG. 3. (a) Parabolic and linear dispersion relations corresponding to metal and graphene electrodes, respectively. At $T = 0$ only states lying in the shaded region contribute to the tunneling current. For a given $k_z$ both the in-plane momentum $k$ and the total energy $E$ are conserved only for the states with in-plane momentum $k_1$ in the CB and $k_2$ in the VB. (b) Work function $W_m$, equilibrium separation $d_{eq}$, short range interaction $\Delta_{ch}$, and $V_D$ voltage as defined in the main text, for several metals under study.

At low temperature ($T \to 0$) the expressions for both $\Delta E$ (Eq. 2) and $J$ (Eq. 8) have a closed-form analytical solution. However, numerical calculations at $T = 300K$ will also be presented in this section. As $T \to 0$ the $f$ function from Eq.

(2) reduces to $f = \pm\Delta E^2/2$, where the upper (lower) sign applies for $\Delta E < 0$ ($\Delta E > 0$) and then Eq. (2) becomes a quadratic equation for $\Delta E$ whose solution is

$$\Delta E = \pm \frac{\sqrt{1+4\alpha e|V-V_D|}-1}{2\alpha}. \quad (10)$$

The upper sign applies for $V \leq V_D$ and the lower sign for $V > V_D$. Fig. 4 shows the behavior of $\Delta E$ as a function of the bias voltage $V$ in MG juntions for different metals with $d = d_{eq}$ and assuming $\varepsilon = \varepsilon_0$ (Fig. 3b). We can observe that at $V = 0$, metals such as Cu and Al dope the graphene n-type while Pt and Au electrodes result in p-type graphene.

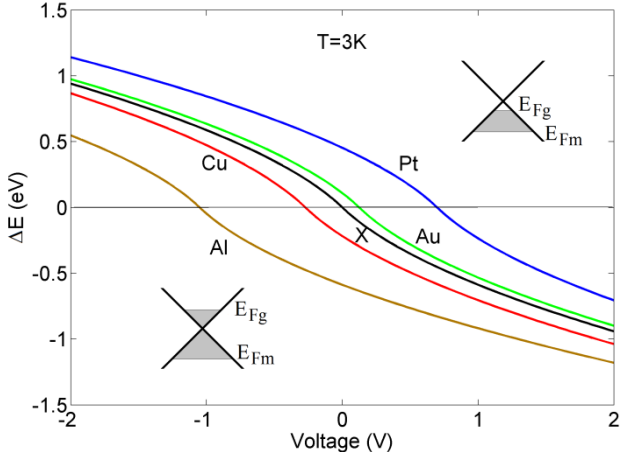

FIG. 4 (Color online). Graphene Fermi level shift with respect to the Dirac point as a function of the bias voltage $V$ for different metals with $d = d_{eq}$.

With the calculated $\Delta E$ given in Eq. (10), an analytical expression for the tunneling current density at $T = 0$ in the MIG structure can be found. Using Eq. (8) we have

$$J = \eta(\kappa)\frac{2m}{\hbar^2}\int_{k_{zmin}}^{k_{zmax}} dk_z \omega(k_z) \int_{\min(k_{min},k_{max})}^{\max(k_{min},k_{max})} kdk\delta[(k-k_1)(k-k_2)], \quad (11)$$

where the meaning of $k_{zmin}$, $k_{zmax}$, $k_1$ and $k_2$ have been graphically represented in Fig. 3. It is worth noting that at $T = 0$ only states with energies $E \in [E_{fm}, E_{fg}]$ contribute to the tunneling current. The energy associated with $k_1$ is normally outside of the energy range for which the graphene dispersion relation is linear ($\sim \pm 1$eV, corresponding to $V \sim \pm 2$V). Both $k_{zmin}$ and $k_{zmax}$ values can be written as

$$k_{zmin} = \sqrt{k_D^2 - 2k_{min}\xi - k_{min}^2}, \quad k_{zmax} = \sqrt{k_D^2 + 2k_{max}\xi - k_{max}^2}, \quad (12)$$

where $k_{min} = (eV + \Delta E)/\hbar v_f$ and $k_{max} = |\Delta E|/\hbar v_f$.

**III. Results.** In this section we report on the tunneling current and differential resistance of the MG junction and MIG heterostructure. Fig. 5 shows, at $T = 300$K, the tunneling current density of the MG junction as a function of $V$ for Pt, Au, Cu and Al metal electrodes, with work functions and equilibrium separation distances given in Fig. 3b, which were taken from Ref. [9]. For the sake of comparison we have also shown the tunneling current for a metal X with work function $W_m = W_g = 4.5$eV, so that graphene is undoped at $V=0$. Although not shown in Fig. 5, the tunneling current at $T = 0$ calculated by Eq. 8 is barely distinguishable from the case $T = 300$K. The main difference is the slope of the curve I-V when $V$ is such that the doping type in graphene changes. Here, we have used the model for $\Delta_{ch}$ given in Ref. [9] to calculate the short range interaction at the equilibrium separation. The $\kappa$ term appearing in the tunneling current expression has the form $\sqrt{(2m\phi/\hbar^2 + k_\parallel^2)}$ [10], where $\phi = 5$eV has been taken as the barrier height, and $k_\parallel$ is the parallel momentum. For graphene, the latter term is essentially equal to the momentum at the K or K' points (i.e., $4\pi/3a$) so that $\kappa = 20.5$nm$^{-1}$. Changes of the concavity of the I-V curves occur whenever $V = V_D$ or $V = V_c$, i.e., when the graphene (Fig. 2b) and metal (Fig. 2e) Fermi levels align with the Dirac point, respectively.

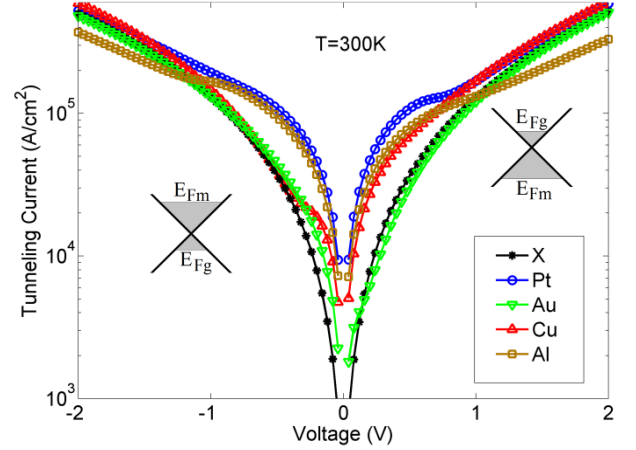

FIG. 5 (Color online).Tunneling current density $J$ for the MG junction as a function of the bias voltage $V$ for different metal electrodes at the equilibrium separation.

Next, we report on the differential contact resistance, defined as $R = G^{-1}$, where $G = dJ/dV$ is the differential contact conductance. An useful expression for the conductance can be found with our model at zero temperature,

$$G = \eta(\kappa)\frac{2m}{\hbar^2}\left[F(k_{zmax}(V))\frac{dk_{zmax}}{dV} - F(k_{zmin}(V))\frac{dk_{zmin}}{dV}\right] \quad (12)$$

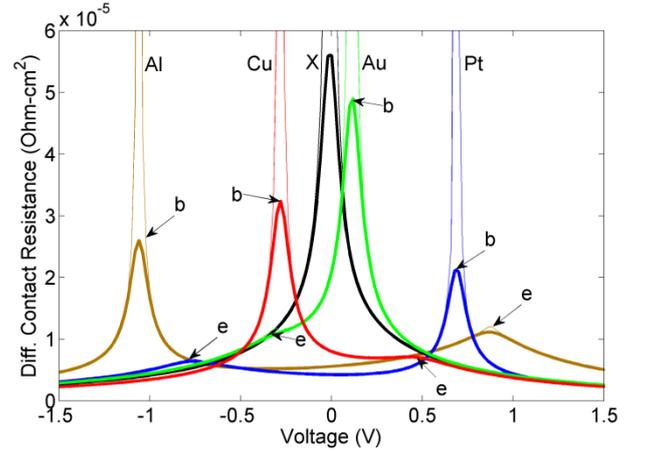

FIG. 6 (Color online). Differential contact resistance of the MG junction as a function of the bias voltage $V$ for different metal electrodes in equilibrium separation. Thick and thin lines are for $T = 0$ and 300K, respectively. The letters "b" and "e" refer to the situations described in Figs. 2b and 2e, respectively.

where the function $F(x) = \omega(x)k_2(x)/|k_2(x) - k_1(x)|$ is the argument of the integral in Eq. (11). Figure 6 shows the differential contact resistance as a function of V, which exhibits "b" points (corresponding to the situation $V = V_D$) for Au and Pt at $V > 0$. For Al and Cu these "b" points are located at $V < 0$. For the MG juntion with X metal, the only peak appears at $V=0$, similar to MIM diodes [12]. The "b" and "e" points refer to the situations in Figs. 2b and 2e, respectively.

Next, we show the behavior of the differential resistance as a function of the (I) layer thickness for the different MIG structures including Pd and Ti (Fig. 7). These latter metals were not previously discused in the context of MG junction theory because the strong graphene-metal bonding interaction destroys the conical points, and our model is no longer valid in this situation. We have assumed a permittivity $\varepsilon = 4\varepsilon_0$ for the (I) layer. Here, the expected exponential dependence with $d$ can be observed. On the other hand, the $V_D$ voltage changes, for every metal, with respect to the MG juntion because now $\Delta_{ch}=0$. For instance, if we consider Cu as a metal and a bias $V=0$, the graphene of MG junction is n-type (Fig. 6) but, in a MIG structure instead, graphene can be p-type. In addition, Fig 7 shows that the voltage for which the "b" points arise is insensitive to the interfacial layer thickness $d$ while the "e" points move with $d$. This is because "b" points only depend on electrical properties while the "e" points dependent on system geometry.

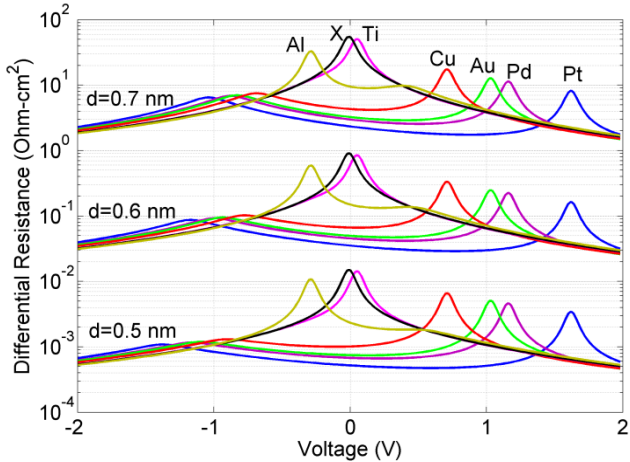

FIG. 7 (Color online). Differential resistance of the MIG heterostructure and its dependence on $d$, where $\varepsilon = 4\varepsilon_0$ and $T = 300K$ have been assumed.

Fig. 8a shows the asymmetry factor, which is defined as the magnitude ratio of the reverse current at $-V$ to forward current at $V$, being a figure of merit for MIM diodes. The asymmetry factor of the MG with Al reaches ~2.2 at 0.2V and in the X diode decreases monotonically, which can be explained by the differences between the dispersion relations of graphene and metals and the difference between the $k_2$ values in the CB and the VB given in Eq. (9). Fig. 8b represents the variation of the differential contact resistance in "b" points as a function of the temperature which is attributed to the reduced DOS in the graphene when $\Delta E = 0$. In contrast, although it is not shown, the resistance in the points "e" do not change significantly.

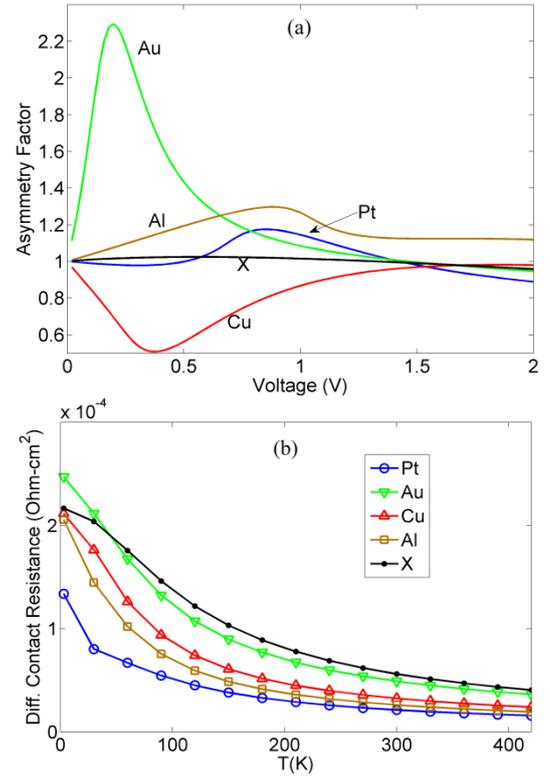

FIG. 8 (Color online). (a) Plot of the asymmetric factor, defined in the main text. (b) Differential contact resistance for the "b" points as a function of the temperature.

In conclusion, we have developed a tunneling current model for the MIG heterostructure based on the BTH method. The model reveals the role played by the electrical and physical parameters in determining the differential resistance. In particular, the role played by the metal work function and insulator thickness has been elucidated.

## ACKNOWLEDGMENTS

We acknowledge the support from SAMSUNG within the *Global Innovation Program*.